# Connections: the relationships between Neolithic and Bronze Age Megalithic Astronomy in Britain
## Gail Higginbottom & Roger Clay

**ABSTRACT:** It has already been empirically verified that for many Bronze Age monuments erected in Scotland between 1400-900 BC, there was a concerted effort on behalf of the builders to align their monuments to astronomical bodies on the horizon. It has also been found that there are two common sets of complex landscape and astronomical patternings, combining specific horizon qualities (like distance and elevation) with the rising and setting points of particular astronomical phenomena. However, it has only been very recently demonstrated by us that that the visible astronomical-landscape variables found at Bronze Age sites on the inner isles and mainland of western Scotland were *first established* nearly *two millennia earlier*, with the erection of the *mooted first standing-stone* 'great circles' in Britain: Callanish and Stenness of Scotland. In *this* paper we demonstrate the connection between all of these monuments and the large LN circles south of Scotland, namely those of Castlerigg and Swinside in Cumbria, England.

**The standing stones of Scotland**
The chronology, archaeological associations and various possible functions of standing stone monuments, discussed at length in Higginbottom *et al.* 2013, reveal a number of informative points on the archaeology of free-standing stones (F–SS) in Scotland.[1] This archaeological information tells us that these monuments seemed to appear suddenly and grandly during the Late Neolithic (LN) approximately 3000-2900 BC and were built until the end of the Bronze Age (BA). The mooted first F-SS built during this time were great circles, using thin, tall slabs: Callanish 1 and Stenness of Scotland.[2] Fascinatingly, *linear* F-SS

---

[1] G. Higginbottom, A.G.K. Smith and Philip Tonner, 'A Re-creation of Visual Engagement and the Revelation of World Views in Bronze Age Scotland ', *The Journal of Archaeological Method and Theory* (2013): doi:10.1007/s10816-013-9182-7;

[2] P. J. Ashmore, *Calanais Survey and Excavation 1979-88* with contributions by T Ballin, S Bohncke, A Fairweather, A Henshall, M Johnson, I Maté, A Sheridan, R Tipping and M Wade Evans. (*in press*).; Patrick Ashmore, 'Radiocarbon dating: Avoiding errors by avoiding mixed samples', *Antiquity* 73 no. 279, (1999): pp. 124–130; Rick Schulting, A. Sheridan, R. Crozier, & E. Murphy, 'Revisiting Quanterness: New AMS dates and stable isotope data from an Orcadian chamber tomb'. *Proceedings of the Society of Antiquaries Scotland 140*, (2010): pp. 1–50.

*sites* in Scotland are so far scientifically dated within 1400–800BC and single stones are associated with, or part of, monuments dated through the LN to the late Bronze Age (BA). Those that stand alone in Scotland are so far dated to the late BA. Such monuments are mostly *directly* associated with death, fire, burial, body transformations and pale/white/shiny stones or pebbles throughout this same timeframe.[3] Some interpretation of the monuments' role(s) could proceed at this point, but this kind of archaeological information can not tell us why people chose to erect these monuments where they did, and does not explain fully why they erected them at all (qv Higginbottom 2013). In this paper, we will limit ourselves to the dominant context, the surrounding 360-degree landscape, in particular the intersection of the land and the sky – the horizon. This visual boundary, the furthest point a person can see, defines and contains what is to be observed from a megalithic site. Through the examination of this context, we can demonstrate the connection of place (or places) and continuity of cosmology over two millennia more firmly than can the accompanying archaeological evidence, though the latter is essential for any full interpretation and comprehension of these generally enigmatic sites.

**Earliest work**
*Orientation studies - testing the distribution of observed horizon declinations indicated by monument alignments*
Meaningful sets of data and statistical analyses were devised for sites with a single or very small number of orientations (like single slab or stone row), to analyse the astronomical potential of the sites contained within western Scotland as a whole and then within separate geographical regions. The fundamental problem is that the horizon elevation function (the relationship between azimuth and elevation, and therefore too the associated declination) *is real and fixed to a specific site,* it is not a probabilistic distribution. Thus, we had to ensure that we were using real declination data for both our expected and observed data sets. To overcome such limits in the study of archaeoastronomy early on in the project, Smith developed the *Horizon* software, which we applied in our previous statistical investigations.

This process allowed us to discover that for the islands of Mull, Coll and

---

[3] Richard Bradley, ed., *The Good Stones: A new investigation of the Clava Cairns*. (Edinburgh: Society of Antiquaries of Scotland, 2000); Paul Duffy, 'Excavations at Dunure Road, Ayrshire: A Bronze Age cist cemetery and standing stone'. *Proceedings of the Society of Antiquaries of Scotland* 137, (2007): p. 53; J. N. G. Ritchie, 'The Stones of Stenness, Orkney', *Proceedings of the Society of Antiquities of Scotland* 107, (1976): pp. 1–60; Rick Schulting *et al.*, 'Revisiting Quanternes', pp. 35–36.

**Connections: the relationships between Neolithic and Bronze Age Megalithic Astronomy in Britain**

Tiree together with 2 sites from North Argyll, as well as Argyll with Lorn, and Islay with Jura, the *distribution of observed horizon declinations* indicated by monument alignments was unlikely to be due to chance factors (Kolmogorov-Smirnov test, rejection of the null hypothesis: Mull $p=0.00817$ (n1=24 sites; n2= 40 declinations), Argyll $p=0.00593$ (n1=21, n2= 44) and Islay $p=0.00105$ (n1=23, n2= 41).[4] We therefore interpreted this outcome to mean that the monument alignments indicating particular declinations *along the horizon* were deliberately chosen by the builders of these monuments. The subsequent investigation of these declinations found statistical support for an interest in the Moon's rising and setting points most close to the MajLS and MinLS both in the southerly and northerly directions, as well as the Sun at the winter solstice (WSol) and areas that flank the midpoint between the solstices. These statistical tests were carried out on groups of sites across the chronological range from Neolithic to the end of the BA. Whilst no statistical support was found for the Sun at the summer solstice (SSol) by region, a small number of sites were oriented in this direction within 2° (approximately nine orientations out of 276). However, it will be demonstrated by the 3D landscapes below that the SSol in the BA was important in ways similar to the great circles of the Late Neolithic.

*3D-landscape reconstruction with astronomical phenomena layer*
Significantly, *we have found two horizon landscape patterns, one which is basically the topographical reverse of the other*. For our detailed regional studies to date, we have found that one or the other surrounds every site. For all the sites on Coll and Tiree (n=6/6), the majority of sites on Mull (n=9/16) and roughly half of the sites studied so far in Argyll (with Lorn; (n= 20)) there is a combination of *usual* visual cues, whether the sites are linear, single slabs, or small circular settings.[5] We called these 'classic sites' as they contained the first pattern we recognised. The usual dominant cues for classic sites are (Figs. 1 & 2a- c): 1. water is usually seen in a southerly direction; 2. a northern horizon is closest, a southern most distant; 3. the northern horizon has a higher general profile or the highest vertical extents in the profile (apparent elevation); the southern horizon has a very distinct dip (concave) or a lower general profile than the northern; 4. the highest areas of the northern and southern horizons often focus around the four ordinal directions of NW, NE, SW and SE; occasionally the highest area is more generally northern if a single mountain or range fills the northern horizon; 5. the highest points of apparent elevation on the horizon profiles are usually made up of distinct mountains or hills; where there is no mountain or hill range, a single hill or higher ground is usually located near or at these compass points. Whilst most sites have relative peaks near all four ordinal points, some have three; 6. the summer and winter solstitial Sun and standstill Moon tend to rise out of and set into these high ranges, hills, or ground. 7. a site most often forms an alignment internally, or with another site, at a lunar or solar orientation (the majority of which fall within the statistically supported declination ranges). For the Moon this is the Major or Minor Lunar Standstill (MajSS or MinSS), and for the Sun it is at the WSol or SSol. A few are aligned N-S.

---

[4] Higginbottom *et al*. 'Gazing at the horizon' (2001): pp 43-50; G. Higginbottom, AGK Smith, Ken Simpson, and R. Clay, 'More than orientation: placing monuments to view the cosmic order', in Amanda-Alice Maravelia (ed) *Ad Astra per Aspera et per Ludum* BAR, International Series S1154 (Oxford: Archaeopress, .2003): pp. 39-52. With typographical corrections for results' table published in the previous reference.

[5] Higginbottom *et al.* 'Re-creation of Visual Engagement' (2013): 47-53; G. Higginbottom, 'The world begins here, the world ends here: construction of observers of time on the inner isles of western Scotland (Mull)' *in preparation*; G. Higginbottom & AGK Smith, 'Intricate vistas: value driven landscapes in prehistoric Argyll', i*n preparation.*

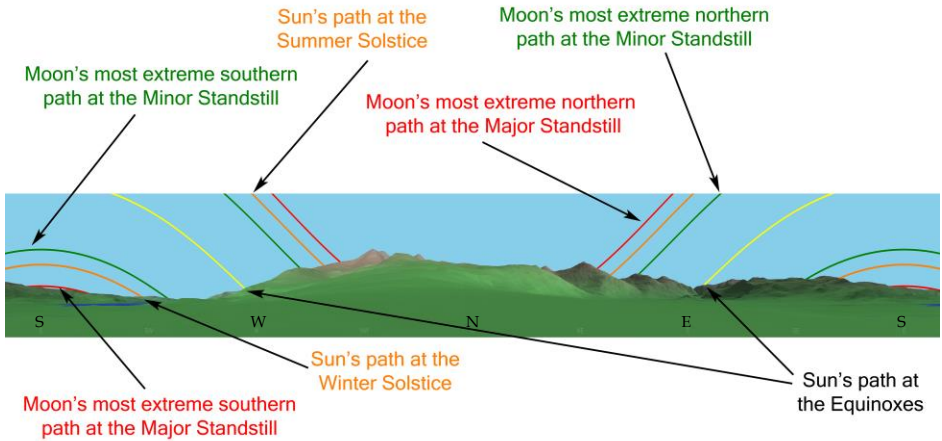

*Fig. 1: This is the 3-D rendering of the landscape around the classic site of Uluvalt on Mull along with key to reading the paths of the sun and the moon on the other such figures below. N=north; S=south. Software created by Andrew Smith. Based upon the Ordnance Survey 1:50 000 Landform PANORAMA map with permission of the Controller of her Majesty's Stationery Office © Crown Copyright.*

As mentioned above, those sites that do not reveal this landscape pattern reveal a combination of reverse landscape traits, namely (see Fig. 2d-e): 1. water is usually seen in the north; 2. a southern horizon is closest, a northern most distant; 3. southern horizon has the highest point(s) in profile; the northern horizon has a very distinct dip or overall lower horizon profile than the4. occasionally the highest area is more generally southern if a single mountain or range fills the southern horizon. 5. It is common for the Moon at the MajSS in the south to be blocked by the horizon (though it is quite clear at a full Moon the glow is visible at particular sites). We call these simply 'reverse sites'. Their remaining astro-horizon qualities remain the same as those found in points 4-7 above. With these astro-landscape patterns, the occasional SSol alignment and a particular astronomical event (where a full moon at the MajSS in the south – the direction *the majority of statistically supported orientations face* - can only occur around the time of a SSol) shows us that the event of the SSol is likely just as firmly entrenched in the consideration of monument placement as those of the statistically indicated alignments. What is important, is that these sites that we have studied so far in these regions of Coll, Tiree, Mull and Argyll, is that they likely to be primarily BA.[6]

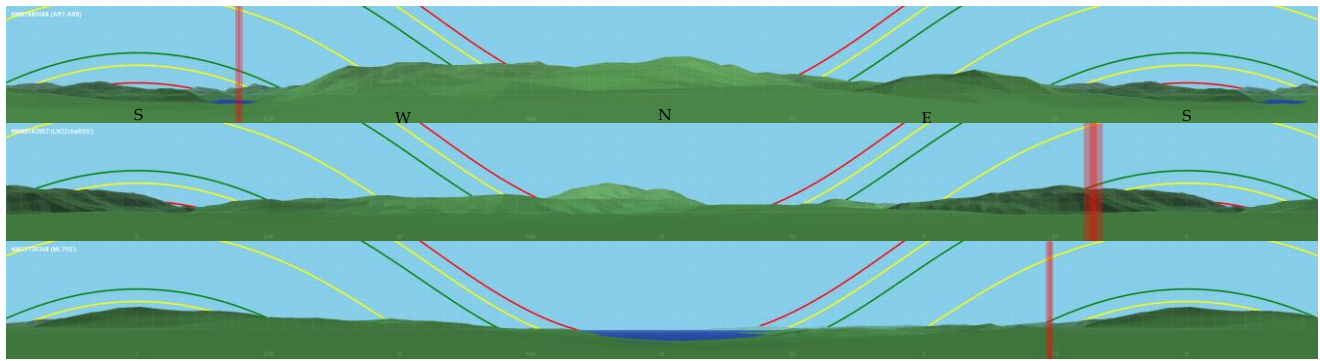

*Fig 2: 3D landscapes of stone rows, slabs and single menhirs, where the centre of the landscape is north. In order from the top: two classic sites of (a) Torran NM87880488 (menhir, AR7 looking towards Ford, AR8), (b) Duachy NM80142052 (stone row, LN22cba) as well as one reverse sits: (c) Cillchriosd (menhir, ML7SE). The red, vertical lines indicate the direction of the alignment of the site and where it touches the horizon. Codes refer to Ruggles' 1984 site numbers. Created with the software Horizon by A. G.K. Smith, ©A.G.K. Smith. Based upon the Ordnance Survey 1:50 000 Landform PANORAMA map with permission of the Controller of her Majesty's Stationery Office © Crown Copyright. Image copyright © Andrew Smith & Gail Higginbottom (2013); http://www.ordnancesurvey.co.uk/docs/licenses/os-opendata-licence.pdf*

**Late Neolithic sites**
*Orientation studies*
Previously, to empirically assess the possible astronomical associations of the mooted earliest standing stones in Scotland which were stone circles (q.v. *Footnote 1*) we designed and applied new methods to formally *test the likelihood* of such connections. Ruggles in his major research project in Scotland chose to dismiss 'from further consideration any on-site indications involving stone rings'.[7] This was because astronomical hypotheses involving sightings across stone rings are dependent upon other variables, such as whether or not the 'site fits a particular geometrical construction' and therefore did not fit with his general research questions at the time. Further, no statistical test had yet been determined to deal with the associated probability issues connected to investigations of looking at orientations within a single circle. Such a test involves separate determinations of the likelihood of various statistical errors, including errors in orientation due to archaeological alignment uncertainties and the uncertainty of which part of the astronomical phenomenon was of interest as it crossed the horizon (e.g. when it first touches the horizon or its final disappearance; thus testing the intentions of the builders).

Another very important issue connected to circles with large numbers of outer stones, is the increased likelihood of hitting an astronomical object by chance, increasing the statistical errors, and therefore reducing the level of probability at which one can reject the null hypothesis. In relation to Ruggles' concern, one can alleviate these by creating a set of *apriori* variables that each site must have to become part of the case-study (see Higginbottom and Clay *submitted* for details about Callanish and Stenness).[8] Regarding the numbers of stones, we now know that circles which have up to **16 stones,** are not high enough to increase the statistical errors significantly, so it is possible to extend the number of stones within a circle to test this further. Finally, we have created a test that can take account of the relevant statistical errors related to alignment uncertainties and the uncertainty of which part of an astronomical phenomenon was of interest as the horizon was crossed as well as including taking into account the number of stones in the circle. See Higginbottom and Clay *submitted*, for details of the statistical tests, of which there is also a brief description below.

---

[6] A. Burl, *From Carnac to Callanish: The Prehistoric Stone Rows and Avenues of Britain, Ireland, and* 530 *Brittany* (New Haven, Yale University Press, 1993). Burl, A. *The Stone Circles of Britain, Ireland and Brittany*. (New Haven, Yale University Press, 2000); Higginbottom *et al.* 'Recreation' (2013): pp. 47-53.
[7] C. Ruggles, *Megalithic astronomy: A new archaeological and statistical study of 300 Western Scottish Sites*. Oxford: British Archaeological Reports British Series 123 (1984): p. 61.
[8] G. Higginbottom, and R. Clay, 'Callanish and Stenness: the Origins of Standing Stone Astronomy in Britain' *submitted to Journal of Archaeological Science*.

The new sites that we incorporate into our study here have been chosen according to age and circle types, ensuring that they shared similarities with Stenness and Callanish for comparison in this paper. Specifically, they should be dated to the LN by scientific dating or by typological association with these dated sites or *long-term archaeological* tradition, their shape should either be geometrically circular like Stenness or a flattened circle like Callanish and their diameter should be greater than 13 m (the size of Callanish).

*The orientation foci of our large Neolithic circles – new work*

We have chosen two sites from the west coast of Britain, this area has a secure megalithic tradition that appears to have lasted for the same amount of time as it did in western Scotland.[9] These southern circles are currently hypothesized to be of possible LN date.[10] Looking at Table 1 we can see that our first new site, Castlerigg (column 3) is a flattened circle and that Swinside (column 4) is geometrically circular. This table informs us that Castlerigg's orientations have 'hit' 10/12 astronomical phenomena (*n*stones=26) and Swinside's have 'hit 8/12 (*n*stones=19).

*Probability analyses of the new great circle's orientations*

For this *we devised* a cross-correlation test which compared the stone directions with the direction of the astronomical phenomena where it crosses the horizon. Specifically, we carried a test to determine the likelihood that the individual megaliths are oriented to the extreme rising and setting points of the sun and the moon - of their annual and metonic cycles, respectively. As a *first step*, we made the assumption that the stone settings and alignments were fully determined and 'surveyed in' with posts, or similar, before any of the megaliths were put in place, and as the stones were of varying widths of up to 7-8 degrees we made a conservative nomination for the orientation error of +/-3 degrees.

For brevity, and thus to fit within the scope of this publication, we have created a table and included the list of possible astronomical phenomena of interest (targets) for the individual stone alignments of each circle and indicated whether: (1) the monument orientation 'hits' any astronomical phenomena within 'total error' ('astronomical error' + 'orientation error' as well as (2) the calculated likelihood (*p*) of the number of 'hits' at each site coming from random chance. We have also included Callanish and Stenness, as well as information on the probable BA monuments of western Scotland for comparison. The results of our tests show us that there is support for both monuments having individual megaliths that are oriented to the extreme rising and setting points of the sun and the moon where the total number of astronomical targets are 8/12 (66%) for Castlerigg and 9/12 (75%) for Swineside. The probability (*p*) that the number of hits is due to chance is 0.0014 and 0.0069, respectively. Strongly supporting that, given the number of stones in the circle, the number of individual hits found is very unlikely to be due to chance. Importantly, Table 1 also demonstrates that there is interest in the same astronomical phenomena at the simpler BA sites and it is clear that there are also similar combinations of astronomical targets where, for example, opposite directions of a single stone row, or a combination of an internal alignment and an alignment with another site, can contain two different lunar alignments; or two parallel monuments side-by-side or close-by might contain a solar and one or more lunar alignments.[11] Relevantly, BA sites can cluster nearby such that a larger number of targets are covered within a small local area, such as in the Kilmartin Valley, Argyll.

---

[9] Burl, *From* Carnac (1993); Burl, *Stone Circles* (2000).
[10] Burl, *From* Carnac (1993); Burl, *Stone Circles* (2000).

*Late Neolithic - 3D-landscape reconstruction with astronomical phenomena layer*
When we observe Fig. 3 of EN stone circles it is striking how similar the horizon shapes, and the positioning of the rising and setting astronomical bodies in relation to this, are to those of the seemingly simpler monuments of the LBA seen in Fig. 2. Whilst these similarities are easy to see we shall point out a few points of interest. Firstly we can see for both sets of figures that whether the overall landscape is mountainous or flat, hills or peaks are chosen in the ordinal directions where possible. For Castlerigg (Fig. 3a), the northern astronomical phenomena rise out of the flanks dominant peaks in the NE and set into the top or the flanks of that in NW, and these peaks are clearly higher in the north than the south. In the south there is the usual dip in the horizon with the phenomena's path rising and setting either-side. Swineside is an example where there is only one major peak in the north being higher than the southern as we have found elsewhere, like on Coll and Tiree, nevertheless, there are still clear

---

[11] Higginbottom 'The world begins here' *in preparation*; C. Ruggles, 'The linear settings of Argyll and Mull', *Archaeoastronomy* (JHA) no. 9 (1985): pp. S105–132.

**Connections: the relationships between Neolithic and Bronze Age Megalithic Astronomy in Britain**

| Site or Region Code | 1 -Circle | 2- Flat Circle | 3- Flat Circle | 4- Circle | 5- SSS, SR, SP | 6 - SSS, SR, SP |
|---|---|---|---|---|---|---|
| *Possible Target* | | | X= HIT within error range | | | |
| NORTH stone or Northern Feature for circles e.g. entrance | X | X | X | X | X | X |
| MinLS rise | | X | Out by 1° | | X | X |
| MinLS set | X | | | X | | X |
| MajLS rise | | X | O | X | X | X |
| MajLS set | X | X | X | | X | X |
| SOUTH stone or Southern Feature | X | X | X | X | X | X |
| MinLS rise | X | | | X | X | X |
| MinLS set | | | X | X | X | X |
| MajLS rise | | X | X | Out by 0.7° | X | X |
| MajLS set | X | X | Xx2; | | | |
| Total lunar totals n=8 | 4/8 | 5/8 | 4/8 (but 5/8 hits) | 4/8 | 7/8 | 8/8 |
| SS rise | X | X | X | X | X | X |
| SS set | | | X | X | X | X |
| WS rise | | X | X | X | X | X |
| WS set | Out by 0.3 | X | Out by 0.5° | | X | X |
| Equinox -rise | Not tested | Not tested | X | X | | |
| Equinox -set | Not tested | Not tested | | Out by 0.7° | | X |
| Total solar totals n=4 or 6 | 1/4 | 3/4 | 4/6 | 4/6 | 4/6 | 5/6 |
| Total astronomical targets | 7/12 | 8/12 | 8/14 | 9/14 | 11/14 | 13/14 |
| Probability (p) that the no of hits is due to chance | 0.0125 | 0.0166 | 0.018 | 0.0083 | N/A | N/A |

*Table 1: Outcomes from our statistical tests for circles and comparisons with regions dominated by stone rows and single standing stones. Site codes are: Scottish sites: (1) = Callanish, Lewis, Scotland (2) Stenness, Orkney; English sites: (3) Castlerigg, Cumbria (note that the "O" for the MajLS rise indicates an internal alignment from the centre through the middle of a cairn within the circle's NE quarter – not included in the statistical test for reasons of continuity; see note re this site on landscape regarding hits for MajLS set), (4) Swinside, Cumbria; and regions from the west Scottish coast primarily made up of single standing stones (SSS), pairs (SP) and rows (SR) (5) Mull, with Coll, Tiree, & Nth Argyll and (6) Argyll, with Lorn. Note these sites are often aligned exactly or very close to N-S also and usually in association with the rising or setting of a southern phenomena. Note possible hits for Castlerigg and Swinside that were beyond our stringent and conservative error range that could not be used by the statistical test. Conservative in that our stone error was often even narrower than the actual stone widths themselves.*

the SE (both much more obvious in a full-size version of the figures; these figures are about 1/10[th] the size of the original size). Again there is the clear dip in the horizon profile in the south and prominent peaks in the SE and SW, into separate hills associated with the rising phenomena in the north and water in which all phenomena set into in the SW and which the MajLS rises out of in the SE.

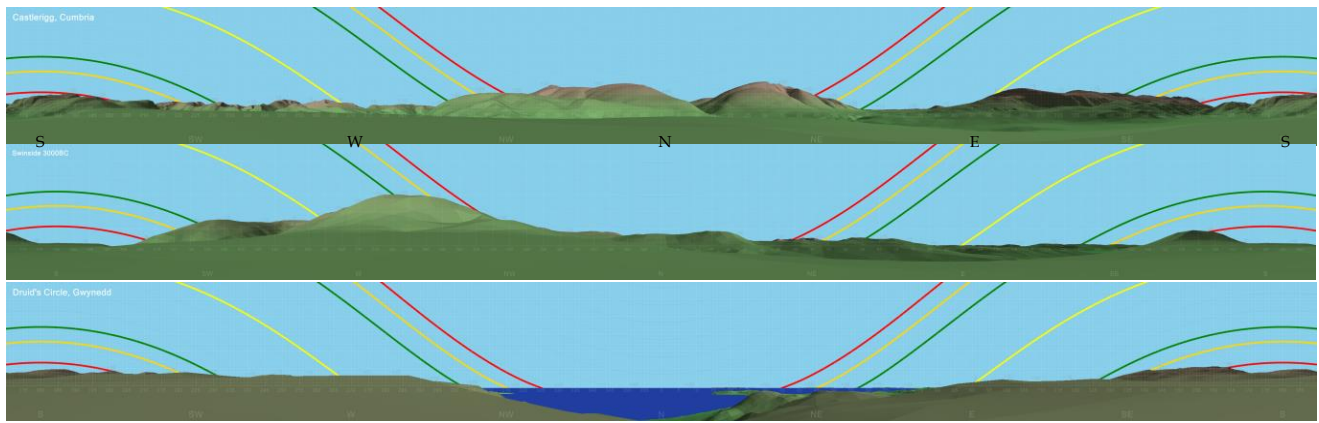

*Fig. 3: 3D landscapes of stone circles. In order from the top: two classic likely Late Neolithic sites of (a) Castlerigg and (b) Swineside, Cumbria, west England; one reverse sites: possible Late Neolithic Druids Circle, Gwenydd, Wales. Created with the software Horizon by A.G.K. Smith, © A.G.K. Smith. Created with Terrain 50. Contains Ordnance Survey data Crown copyright and database right (2012). http://www.ordnancesurvey.co.uk/docs/licenses/os-opendata-licence.pdf. Images copyright © Andrew Smith and Gail Higginbottom (2013).*

The MinLS and WS actually rise out of small hills but this is not evident in such a small figure here. In the SW there is an amazing view of the moon rolling along the horizon for about 12°. In actual fact, the moon actually partially sets and moves partially behind the horizon for about 5 degrees and then fully reappears for 2° and then almost disappears completely again before reappearing fully and finally disappearing for the last time shortly after that. What is extraordinary, is that there is a stone aligned with each of the three times the moon

**Connections: the relationships between Neolithic and Bronze Age Megalithic Astronomy in Britain (draft)**

hits the horizon to start setting. Our studies of EN stone circles sites in western Britain is ongoing and we have added an example of reverse site circle further south again on the western side of Britain with similar megalithic traditions.

## Discussion and Conclusions

Their landscape and astronomical choices in relation to horizon distances and direction, horizon profile (mountain/hill locations, lower and higher ground) and the astronomical phenomena associated with each of these - are clearly consistent between the Early Neolithic and Late Bronze Age sites also.

Therefore our statistical results, along with the 3D landscape formations, further support our initial discovery that early stone circles and the later Bronze Age monuments of western Scotland share building and knowledge traditions which incorporate their interest and knowledge of astronomical. It has not yet been proven that early and mid-Bronze Age monuments share exactly the same traditions as yet. Further more, more sites must be examined and excavations are rare carried out to acquire dating evidence of the likely earliest circles. This will provide more secure dating evidence like that found for Callanish. We look forward to continuing making important discoveries in our *Western Scotland Megalithic Landscape Project*.

## Acknowledgments to date

We would like to thank Andrew Smith for the use of his software *Horizon* and for all the advice attached to this. We would also like to thank Fabio Silva (an editor of the proposed volume) for his constant technical assistance and answering of queries. Significantly, we thank and acknowledge the European Society for Astronomy in Culture (SEAC) committee for *the opportunity of submitting this paper* to the referred proceedings of SEAC 2014: The Materiality of the Sky, co-hosted by the University of Malta and Heritage Malta. We are currently awaiting reviewers' comments.

Gail Higginbottom and Roger Clay